\makeatother

\documentclass{ifacconf}

\usepackage{mathrsfs,amsmath,amssymb,amsfonts,mathtools}
\usepackage{color}

\usepackage{graphicx}      

\usepackage{natbib}        

\usepackage{algorithm}
\usepackage{algpseudocode}

\newtheorem{example}{Example}

\begin{document}
\begin{frontmatter}

\title{
Accelerating reaction network identification via frequency-domain analysis\thanksref{footnoteinfo}} 

\thanks[footnoteinfo]{
This work was supported by the National Natural Science Foundation of China under Grant No.12501699 (to Z.F.) and No. 62303409 (to X.Z.), and the Jiangsu Provincial Scientific Research Center of Applied Mathematics under Grant No. BK20233002 (to X.Z.).}

\author[First]{Zhou Fang} 
\author[Second]{Wenying Sun}
\author[Second]{Xiaoyu Zhang}
\author[forth]{Mustafa Khammash}

\address[First]{State Key Laboratory of Mathematical Sciences, Academy of Mathematics and Systems Science, Chinese Academy of
Sciences, Beijing, 100190, China. (e-mail: zhfang@amss.ac.cn)}

\address[Second]{School of Mathematics, Southeast University, Nanjing, 211189, China. (e-mail: xiaoyu\_z@seu.edu.cn)}

\address[forth]{Department of Biosystems Science and Engineering, ETH Zurich, Basel, Switzerland. (mustafa.khammash@bsse.ethz.ch)}

\begin{abstract}                %
    The identification of intracellular reaction networks from single-cell data lies at the heart of many biological studies, as it can unravel key mechanisms in living organisms and provide insights for their rational engineering.
    However, the stochastic and nonlinear nature of these reacting systems poses significant challenges for this identification problem, making the state-of-the-art methods (e.g., particle-filtering-based approaches) computationally demanding.  
    This paper reports a noteworthy numerical observation that for certain systems, the Fourier transform can render the likelihood of the measurement data approximately Gaussian, therefore resulting in a substantially accelerated Bayesian inference algorithm. 
    A gene transcription example is presented to illustrate this fin  ding and demonstrate the efficiency of {the proposed Fourier-based method}.
\end{abstract}

\begin{keyword}
Reaction networks, 
system identification, 
frequency-domain analysis,
Bayesian inference.
\end{keyword}

\end{frontmatter}

\section{Introduction}

Modern biotechnologies have enabled scientists to track intracellular reaction processes at high temporal resolution \citep{stephens2003light}. 
Identifying the dynamic model of such intracellular reaction systems from these single-cell time-course data lies at the heart of many biological studies, as it provides crucial insights into living systems and lies the foundation for their control. 
Despite its importance, this identification problem remains challenging due to the stochastic and nonlinear nature of these systems \citep{browning2020identifiability, schnoerr2017approximation,bronstein2015bayesian}.
Notably, these features suggest that the likelihood function---a key component in Bayesian statistics---often has no closed-form expression, thereby posing significant challenges for the statistical inference of the underlying system.

Many efforts have been devoted to this problem by employing stochastic simulations to approximate the likelihood and then combining them with established Bayesian computational methods for system identification. 
Representative approaches include the well-known particle Markov Chain Monte Carlo \citep{andrieu2010particle,golightly2011bayesian,browning2020identifiability}, approximate Bayesian computation \citep{toni2009approximate}, among others.
Although these methods have achieved numerous successful applications in biology, they still face significant computational challenges, as the likelihood approximation requires a large number of Monte-Carlo simulations. 
This difficulty arises because Monte-Carlo methods are generally inefficient in estimating rare events, and the likelihood essentially represents the rare probability of observing the given data.
While techniques such as the control variates \citep{backenkohler2022variance}, marginalization methods \citep{zechner2014scalable,fang2024advanced}, and importance sampling \citep{golightly2019efficient,ben2023learning,wildner2019moment} can mitigate this issue to some extent, they still require substantial computational resources.
Overall, the existing simulation-based likelihood approximation methods might not be optimal for reaction network identification. 

In contrast to directly computing the likelihood, certain data transformations can ease system identification 
\citep{gupta2022frequency,messenger2024weak,heitzman2025practical}.
In particular, researchers have found \citep{gupta2022frequency,song2019frequency,jia2021frequency} that the power spectral density (a Fourier-based transformation of time-series data) encodes rich information about the underlying system, thereby opening a new door for system identification.
Since the empirical power spectral density converges to its theoretical counterpart in the limit of infinitely long data, one can, in principle, infer the system by comparing these two parts when sufficiently long data is provided (see the references above). 
However, this convergence is particularly slow, and the spectral estimates become highly noisy for data of moderate length, which undermines this direct comparison method.

Nevertheless, these studies of power spectral density demonstrate that the frequency-domain analysis, one of the most fundamental concepts in control theory, has the potential for more efficient Bayesian inference of dynamical systems. 

This paper follows this idea and demonstrates that the frequency-domain analysis {might be able} to significantly accelerate likelihood approximation, leading to a more efficient identification method.
Specifically, we show by numerical simulations that for certain reaction networks, the Fourier transform can render the likelihood of the transformed data approximately Gaussian.
This Gaussianity, in turn, suggests that accurate estimation of the mean and covariance matrix suffices for likelihood computation, thereby substantially reducing the required number of Monte-Carlo simulations and improving algorithmic efficiency.
We also present a biologically relevant numerical example to illustrate the superior performance of this {Fourier-based method}.

\section{Identification for stochastic reaction networks}

We consider a reaction network involving $r$ reactions
\begin{align*}
    \nu_{1,j} S_1  + \dots + \nu_{n, j}S_n
    \to \nu'_{1,j} S_1 + \dots + \nu'_{n, j}S_n && j=1,\dots, r
\end{align*}
where $S_1, \dots, S_n$ represent $n$ different chemical species, and $\nu_{i,j}$ and $\nu'_{i,j}$ are stoichiometric coefficients indicating the species changes in each reaction.
When the species have low copy numbers, this reaction system is often modeled by a continuous-time Markov chain \citep{anderson2015stochastic}
\begin{align}\label{eq. reaction network}
    X(t) = X(0) + \sum_{j=1}^r \zeta_j R_j\left(t \right)
\end{align}
with $X(t)$ an $n$-dimensional vector indicating the copy numbers of each species at time $t$, and $\zeta_j \triangleq (\nu'_{1,j} - \nu_{1,j},\dots, \nu'_{n,j} - \nu_{n,j})^{\top}$.
Here, $\{R_{j}(t)\}_{j=1,\dots,r}$ are counting processes that increase by one at the rates of $\lambda_j(\Theta, X(t))$ with $\Theta$ unknown system parameters. 
Also, the immediate increments of these counting processes are conditionally independent given $(\Theta, X(t))$. 

Often, the system parameters $\Theta$ are unknown and need to be inferred from time-course measurements of this dynamical system. 
We consider that system \eqref{eq. reaction network} is measured sequentially at evenly spaced time points $\{t_1, \dots, t_m\}$ and the readouts (denoted by $Y(t)$) satisfy 
\begin{align}\label{eq. system observations}
    Y(t_i) = h(\Theta, X(t_i)) + \sigma W_i
\end{align}
with $h(\cdot)$ a known function, $\sigma$ the observation noise intensity, and $\{W_i\}_{i=1,2,\dots,m}$ standard Gaussian noise. 
For simplicity, this paper assumes that $Y(t_i)$ are scalars; however, the result can be easily extended to higher-dimensional cases.
By denoting $D\triangleq \left(Y(t_1), \dots, Y(t_{m})\right)^{\top}$, the Bayesian inference of parameters $\Theta$ aims to compute the posterior probability density
\begin{align}\label{eq. posterior}
    p(\theta| D) ~\propto~  L(D|\theta) \, p_0(\theta)
\end{align}
where $L(D|\theta)$ is the likelihood of obtaining the data $D$ given parameters $\Theta = \theta$, and $p_0$ is the prior distribution of $\Theta$ representing the initial guess of these parameters. 
This posterior reflects people's belief about the unknown parameters $\Theta$ given the observation data. 

\begin{example}[Running example]\label{ex running example}
Throughout the paper, we consider a telegraph model
\begin{align*}
    & G \rightarrow G^*  && \text{(gene activation)}\\
    & G^* \rightarrow G && \text{(gene deactivation)}\\
    & \emptyset \rightarrow \text{mRNA}
    && \text{(transcription)}\\
    & \text{mRNA} \rightarrow \emptyset
    && \text{(degradation)}
\end{align*}
where $G$ stands for inactivate gene state and $G^*$ represents the active one. 
The system state is denoted by a 3-dimensional vector $X(t) = \left(X_1(t), X_2(t), X_3(t)\right)^{\top}$, representing the copy numbers of $G$, $G^*$, and mRNA, respectively, and it also contain two unknown system parameters $\Theta=(\Theta_1,\Theta_2)$ corresponding to gene activation and deactivation processes. 
The propensity functions are modeled by $\lambda_\text{act}(\theta, x) = \theta_1 x_1$, $\lambda_\text{deact}(\theta, x) = \theta_2 x_2$, $\lambda_\text{tran}(\theta, x) = 30\,x_2 $, and $\lambda_\text{deg}(\theta,x) = x_3$ for the gene activation, deactivation, transcription, and degradation reactions. 
We assume that these unknown system parameters have a uniform distribution in the region $[0.1, 1] \times [0.1, 1]$, and the initial condition is $X(0)= (1, 0, 0)$.
Moreover, we consider that the mRNA copy number (i.e., $X_3(t)$) is measured every 1 minute for a period of 240 minutes, with the readout satisfying 
\begin{align*}
    Y(t_i) = X_3(t_i) + 0.1W_i    
\end{align*}
with $t_i = i$.
Then, the aim is to infer the unknown system parameters $\theta_1$ and $\theta_2$ from the readouts $D \triangleq  \left\{Y(t_1),\dots, Y(t_{240})\right\}$ with the Bayesian framework.
\end{example}

\subsection{Challenges in computing likelihood}

One key challenge in the inference problem \eqref{eq. posterior} is the computation of the likelihood $L(D|\theta)$.
This likelihood $L(D|\theta)$ is often analytically intractable due to the non-linearity of the dynamical system \eqref{eq. reaction network} in most practical applications. 
In this case, the state-of-art approach for computing $L(D|\theta)$ is utilizing sampling-based methods according to the rule  
\begin{align*}
    L(D|\theta) &= \prod_{i=1}^m p_{\theta}\left(Y(t_i)\big| Y(t_1),\dots, Y(t_{i-1})\right)\\
    & = 
    \prod_{i=1}^m \Bigg[\sum_{x} 
    p_{\theta}\left(Y(t_i) \big| X(t_i)=x\right)\\
    &\qquad\qquad\quad\times
     p_{\theta}\left(X(t_i)=x\big| Y(t_1),\dots, Y(t_{i-1})\right)
    \Bigg]
\end{align*}
where $p_{\theta}\left(Y(t_i)|Y(t_1),\dots, Y(t_{i-1})\right)$ is the conditional density of the measurement at time $t_i$ given the parameters $\Theta= \theta$ and all previous observations, 
$p_{\theta}\left(Y(t_k) | X(t_k)=x\right)$ is the likelihood of this measurement give the latent state  $X(t_k)=x$, 
and $p_{\theta}\left(X(t_i)=\cdot~| Y(t_1),\dots, Y(t_{i-1})\right)$ is the predictive distribution of the hidden state $X(t_i)$ given the  parameters $\theta$ and all previous observations.
Based on this relation, one can approximate the likelihood $L(D|\theta)$ using a particle filter \cite{andrieu2010particle,golightly2011bayesian,browning2020identifiability})
\begin{align}\label{eq. particle filtering based computation}
     L(D|\theta) \approx \prod_{i=1}^m
    \left[
        \frac{1}{N}\sum_{\ell = 1}^N
        p_{\theta}\left(Y(t_i) \big| X(t_i)=\tilde X^\ell(t_i)\right)
    \right]
\end{align}
where $\tilde X^1(t_i),\dots, \tilde X^{N}(t_i)$ (for each $i$) are samples generated by the particle filter to approximate the predictive distribution $p_{\theta}\left(X(t_i)=\cdot~\big| Y(t_1),\dots, Y(t_{i-1})\right)$.
Detailed descriptions of particle filtering (also known as \textit{Sequential Importance Resampling} SIR) can be found in \citep{gordon1993novel,doucet2009tutorial}.

In most cases, the above particle filtering-based approximation is computationally intensive, typically requiring thousands of samples ($\sim10^3$) to accurately approximate conditional distributions even in low- to moderate-dimensional systems \citep{gordon1993novel,doucet2009tutorial,andrieu2010particle,toni2009approximate,fang2022stochastic,fang2023convergence}.
The computational burden becomes even more pronounced in high-dimensional systems or when the observations are highly informative, as this often leads to sample degeneracy and necessitates significantly more Monte-Carlo samples ($\gtrsim10^5$ samples; see \cite{fang2024advanced} for an example) to achieve acceptable approximation accuracy.
When particle filtering is applied to likelihood computation, the required number of samples should be further scaled by the length of observation data to control the relative variance of the likelihood ratio estimator \citep{deligiannidis2018correlated}.
Since the likelihood needs to be repeatedly computed in most Bayesian inference methods---such as Markov Chain Monte Carlo \citep{metropolis1953equation,hastings1970monte,andrieu2010particle}, Sequential Monte Carlo \citep{sisson2007sequential,chopin2013smc2}, and Bayesian deep learning \citep{wang2020survey}---this computational inefficiency poses significant challenges for systems identification.

\section{Accelerating likelihood computation via Fourier transform}

Even though the likelihood $L(D|\theta)$ is often analytically intractable, suitable data transformations can render the likelihood of the transformed data structured and tractable, therefore accelerating the likelihood computation and system identification.
The existence of such transformations has long been established in probability theory \citep{angus1994probability}, and the construction of such transformations has been extensively explored in the machine learning field (i.e., the generative models \cite{goodfellow2020generative,kobyzev2020normalizing}).

\begin{figure}[tbp]
    \centering
    \includegraphics[width=\linewidth]{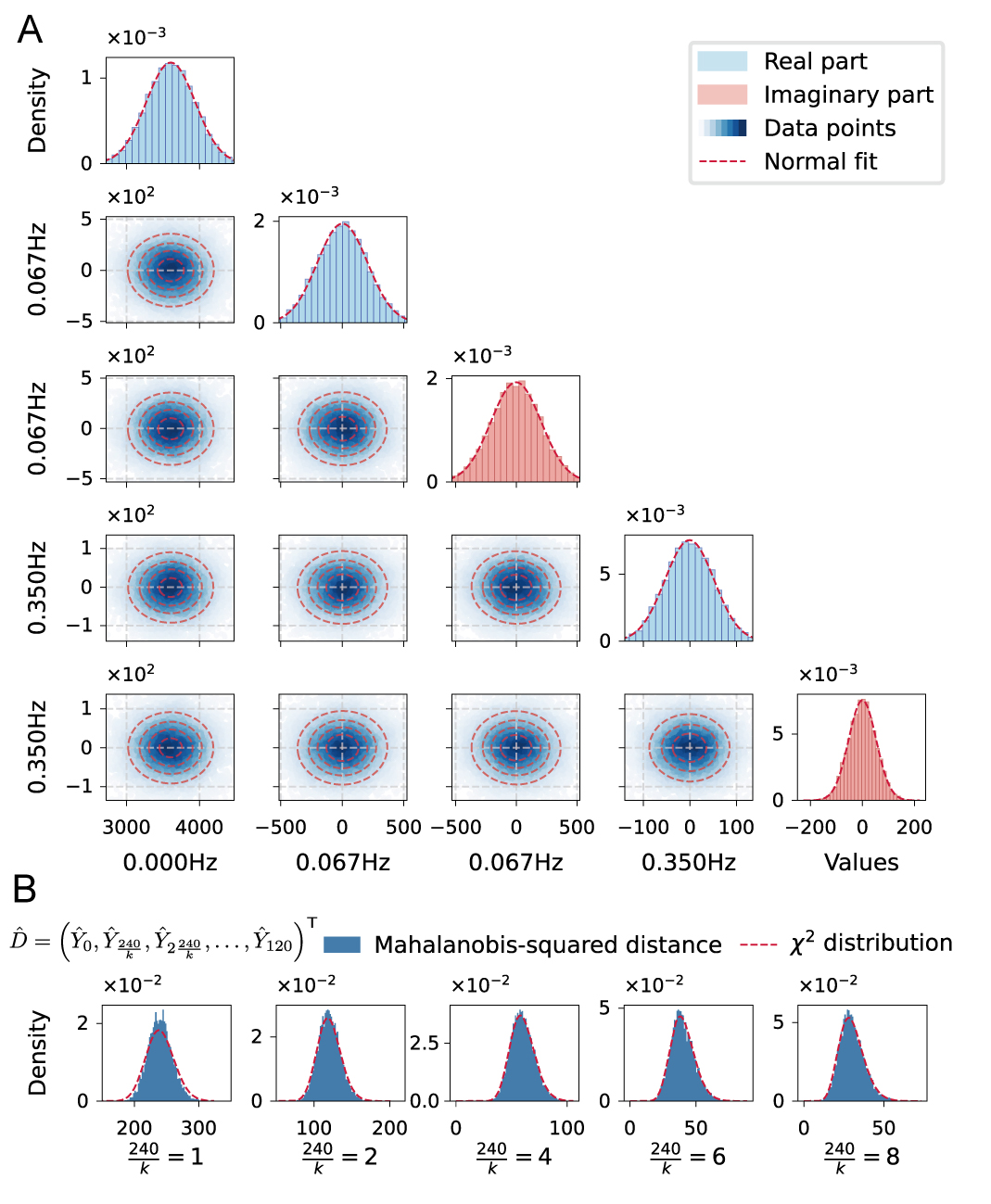}
    \caption{
    \textbf{Test Gaussianity in the transformed obsrvation data $\hat D$ in Example \ref{ex running example}.}
    \textbf{A} Gaussian-like distribution of the transformed observation data $\hat D$ in Example \ref{ex running example}. 
    We generate 10,000 simulations of the system in Example \ref{ex running example} with parameters $\Theta=(0.5, 0.5)$ and compute the Fast Fourier Transform of their observation data. 
    The diagonal panels show the marginal distribution of the transformed data at frequencies 0.000Hz, 0.067Hz, and 0.350Hz, and the off-diagonal panels show their joint distributions. 
    These results demonstrate that given parameter $\Theta$, the transformed data $\hat D$ follows a Gaussian-like distribution.
    \textbf{B} {Gaussianity test} of $\hat D$ encompassing different numbers of selected frequencies.
    This panel shows the distribution of $\hat D$'s Mahalanobis-squared distance, when $240/k$=1, 2, 4, 6, and 8. 
    {Note that} the distribution of {this} distance follows a $\chi^2$-distribution if {$\hat D$ are jointly Gaussian}. 
    The results show that $\hat D$ encompassing every frequency is a bit deviated from the Gaussian distribution.
    {In contrast, when $240/k\geq 4$, the Mahalanobis-squared distance closely approximate a $\chi^2$-distribution, which no longer rules out the possibility that $\hat D$ is approximately Gaussian in these cases.}
    }
    

    \label{fig:normalfit}
\end{figure}

Notably, for certain reaction networks (e.g., Example \ref{ex running example}), the Fourier transform can render the observation data approximately Gaussian.
Specifically, we denote the Fourier transform of the data $D$ by
\begin{align*}
    \hat Y(s) = \sum_{i = 1}^m Y(t_i) \, \text{e}^{-\jmath 2 \pi s  t_i}
    && \text{with $\jmath = \sqrt{-1}$}
\end{align*}
and term $\hat D = \big(\hat Y(s_1), \dots, \hat Y(s_{\ell})\big)^{\top}$ for selected frequencies $s_1, \dots, s_{\ell}$. 
Then, through simulations, we observe that given system parameters, the transformed data $\hat D$ exhibits Gaussian-like distributions (Figure \ref{fig:normalfit}.A); in other words, the likelihood of the transformed data, denoted by $  \hat L\big(\hat D | \theta\big)$, is close to Gaussian. 
Moreover, the transformed data at any two different frequencies turns out to be independent (Figure \ref{fig:normalfit}.A). 
We conjecture that this Gaussianity is a consequence of a central limit theorem for the underlying ergodic system. 
Particularly, at zero frequency, the ergodicity condition guarantees that $\hat Y(0)/m$ converges to the stationary mean and satisfies a central limit theorem with $\frac{1}{\sqrt{m}}\left(\hat Y(0) - \mathbb E\left[\hat Y(0)\right]\right)$ converging to a centered Gaussian distribution \citep{asmussen2007stochastic}.
Owing to the page limit, the full theoretical proof of this Gaussianity will be presented in a forthcoming paper. 

This approximate Gaussianity then allows for a significantly more efficient inference scheme based on the Fourier-transformed data.
Specifically, we can perform Bayesian inference on the transformed data $\hat D$ by expressing the posterior as
\begin{align}\label{eq. posterior from transformed data}
    p(\theta| \hat D) ~\propto~  \underbrace{ \hat L\big(\hat D|\theta\big)}_{
    \tiny\text{Gaussian-like}
    } p_0(\theta)
\end{align}
where $p(\theta| \hat D)$ reflects our belief about the unknown parameters $\Theta$ conditioned on the transformed data $\hat D$.
Then, {the Fourier-based method} utilizes the Gaussianity to efficiently compute the new likelihood $ \hat L\big(\hat D|\theta\big)$; see Algorithm \ref{alg transformed likelihood}.
This algorithm only computes the mean and variance of the distribution $\hat L\big(\cdot | \theta\big)$, and thanks to the Gaussianity, these estimates are sufficient for estimating the likelihood $\hat L\big(\hat D | \theta\big)$.
In contrast to the particle-filtering-based approach \eqref{eq. particle filtering based computation}, which often requires a large amount of samples, the mean and variance estimates of $\hat L\big(\cdot | \theta\big)$ require a much reduced sample size, and, therefore, Algorithm \ref{alg transformed likelihood} has more advanced computational efficiency.
Moreover, this sample size required by Algorithm \ref{alg transformed likelihood} is independent of the observation noise intensity, which makes it particularly advantageous for systems with highly informative observations.
Finally, {our Fourier-based method} integrates this likelihood computation algorithm into existing Bayesian inference frameworks (e.g., Markov Chain Monte Carlo, Sequential Monte Carlo, and Bayesian deep learning) for reaction network identification.

\begin{algorithm}
    \caption{Computation of the new likelihood $ \hat L\big(\hat D|\theta\big)$}
    \label{alg transformed likelihood}
    \begin{algorithmic}[1] 
        \State \textbf{Input:} the parameters $\theta$, simulation sample size $N$, and transformed data $\hat D\triangleq \big(\hat Y(s_1), \dots, \hat Y(s_{\ell})\big)$.
        \State Simulate the system \eqref{eq. reaction network}-\eqref{eq. system observations} with $\Theta=\theta$ to get $N$ simulated observation data $ \mathscr D_1, \dots, \mathscr D_N$ (with $\mathscr D_i = (y(t_1), \dots, y(t_m))$).
        \State Apply the fast Fourier transform to each $\mathscr D_i$ to get 
        $\hat{\mathscr D}_i$.
        \State Compute the mean and diagonal covariance matrix of $\hat{\mathscr D}_1, \dots, \hat{\mathscr D}_N$, and denote them by $\mu$ and $\Sigma$.
        \State Approximate the likelihood $ \hat L\big(\hat D|\theta\big)$ by $\mathcal N\left(\hat D; \mu, {\Sigma}\right)$, where $\mathcal N\left(~\cdot~ ; \mu, {\Sigma}\right)$ is the density of the Gaussian distribution with mean $\mu$ and covariance matrix ${\Sigma}
        $.
    \end{algorithmic}
\end{algorithm}

\subsection{Balancing computational cost and information loss}
\label{section information loss}

While enabling a more efficient inference method, the Fourier-transformed data $\hat D$ can contain less information compared with the original data $D$, when the number of selected frequencies is less than the length of $D$ (i.e., when the data transform is not invertible).
In fact, this information loss can occur because the data $\hat D$ encompassing every frequency from the discrete-time Fourier transform might not be Gaussian (see Figure \ref{fig:normalfit}.B). 
Consequently, the posterior $p(\theta| \hat D) $ based on the transformed data can be less confident than $p(\theta| D)$ in the sense of being more dispersed. 

Despite the information loss, {the proposed Fourier-based method} can still remain advantageous in practical applications, especially when the observed data $D$ is highly informative. 
Recall that in this case, filtering-based approaches are computationally prohibitive, potentially making the result significantly deviated from the target distribution $p(\theta| D)$ and resulting in misleading inferences within feasible computational budgets. 
In contrast, {our method} can efficiently generate accurate approximations to the distribution $p(\theta| \hat D)$, which is typically not too much more dispersed than $p(\theta| D)$, thereby yielding more reliable results. 
{This point is further demonstrated in the numerical example.}

\section{Numerical study of the gene transcription model}

We demonstrate our approach via the telegraph model in Example \ref{ex running example}. {All the numerical experiments were performed on a computing cluster at the Academy of Mathematics and Systems Science, Chinese Academy of Sciences.}

\begin{figure}[tbp]
    \centering
    \includegraphics[width=\linewidth]{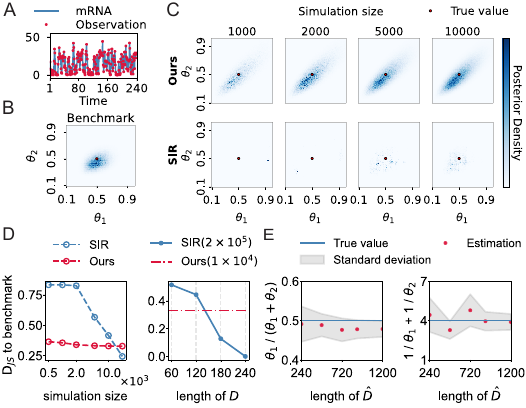}
    \caption{\textbf{Results of the Fourier-based methods applied to the telegraph model.}
    \textbf{A} Simulation of the telegraph model described in Example \ref{ex running example} with parameters $\Theta=(0.5, 0.5)$.
    \textbf{B} Benchmark in estimating the posterior density, {using SIR with a particle size of $200,000$}.
    \textbf{C} Performance of our method and SIR with simulation sizes of 1000, 2000, 5000, and 10000. 
    \textbf{D} {Comparison of the two approaches under different settings.}
    {The vertical axis depicts the Jensen-Shannon Divergence ($\mathrm{D}_{JS}$) between the two approaches and the benchmark.}
    {The first block compares performance under different simulation sizes.}
    {The second block fixes the simulation size (as demonstrated in the legend) and shows the convergence of the SIR approach to the benchmark with the length of the original data $D$; the red line shows our method utilizing the transformed data $\hat D$.}
    \textbf{E} {Parameter estimates from our method over time.
    In these numerical experiments, we set the sample size to be $10^4$ and selected the frequencies $0, \frac{4}{m}, \frac{8}{m}, \dots, \frac{m/2}{m}$ with $m$ the length of the observation $D$.
    The term $\theta_1/(\theta_1+\theta_2)$ represents the fraction of time the gene is activated, and  $1/\theta_1+1/\theta_2$ stands for the average gene on-off circle period.
    }
    } 
    \label{fig:Telegraph_model}
\end{figure}

First, we examined the accuracy of our Fourier-based method in identifying model parameters through numerical simulation and compared the inference results with the SIR method. 
{We set the target system to have parameters $\Theta=(0.5, 0.5)$ and generated an observation trajectory (see Figure \ref{fig:normalfit}.A).}
{For the inference task, we constructed a grid over the parameter space $[0.1, 1] \times [0.1, 1]$ using a $0.01$ mesh and applied the Likelihood-approximation methods to each parameter on the grid to compute the posterior density.}
For our method, we selected the frequencies $0, \frac{1}{60}, \frac{2}{60}, \dots, \frac{30}{60} $. 
Moreover, to get an accurate approximation of the {actual} posterior density, we constructed a benchmark by applying the SIR-based method {with a particle size of 200,000}; the {total} computation time was approximately 15 hours {on} 100 CPU cores. 
{All the results are presented in Figure \ref{fig:Telegraph_model}.}
In both the particle-filtering and Fourier-based approaches, most of the computational effort was spent on simulating the underlying dynamical system. Consequently, the computational cost is effectively determined by the size of the simulation samples; thus, when both methods use the same sample size, they have approximately the same computational cost/time.

{The results in Figure \ref{fig:Telegraph_model} demonstrate that our method is significantly more computationally efficient, requiring only thousands of simulation samples for accurate approximation of the $p (\theta|\hat D)$ (see Figure \ref{fig:Telegraph_model}.C).
In contrast, the SIR-based approach cannot provide reasonable estimates with simulation samples of fewer than ten thousand.
Particularly, when the sample size is less than 2000, the SIR-based approach provides delta distributions, significantly deviating from the benchmark. 
This improvement of our method is attributed to the approximate Gaussianity of the transformed likelihood $\hat L(\cdot|\theta)$, for which accurate estimation of the mean and variance is sufficient for approximation, thereby significantly reducing the required number of Monte-Carlo samples.
}

{Moreover, this efficiency makes our method more advantageous, despite the information loss of the transformed data $\hat D$.
As discussed in Section \ref{section information loss}, data $\hat D$ do not contain all the frequencies and, therefore, has less information than the original measurement data.
This information loss causes the posterior $p (\theta|\hat D)$ more dispersed than $p(\theta|D)$ (see Figure \ref{fig:Telegraph_model}.B and Figure \ref{fig:Telegraph_model}.C).
From the perspective of the distance to the benchmark distribution, the transformed data $\hat D$ has approximately the same information as the first half of the data in $D$ (see Figure \ref{fig:Telegraph_model}.D).
{Also, the distribution $p (\theta|\hat D)$ exhibits a stronger correlation between the two parameter estimates, whereas the benchmark distribution does not.}
Despite this loss, our method is still more accurate than the SIR-based method when the simulation size is kept within a reasonable range (under 10,000), due to the Gaussianity of the transformed data (see Figure \ref{fig:Telegraph_model}.D).
} 

{Furthermore, as the measurement duration increases and more frequencies are incorporated into the transformed data, our method yields sharper parameter estimates, indicated by the narrower confidence interval (see Figure \ref{fig:Telegraph_model}.E).
The numerical results demonstrate that as the data size increases, our method delivers more confident estimates of the fraction of time the gene is activated (i.e., $\theta_1/(\theta_1+\theta_2)$) 
and the average gene on-off circle period (i.e., $1/\theta_1+1/\theta_2$).
}

\begin{figure}[tbp]
    \centering
    \includegraphics[width=\linewidth]{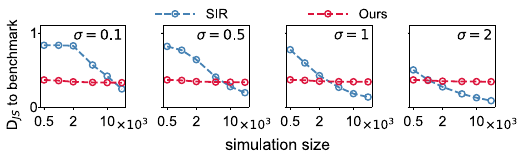}
    \caption{\textbf{Robustness of the two approaches to measurement noise intensity ($\sigma$).}
    We calculate $\mathrm{D}_{JS}$ between the two approaches and the benchmark for various measurement noise levels ($\sigma = 0.1$, 0.5, 1, 2). 
    }
    \label{fig:Sigma}
\end{figure}

To further {demonstrate} the benefits of our method, we investigated the {robustness of our approach to measurement noise intensities $\sigma$} (see Figure \ref{fig:Sigma}). 
The results show that our method {performs consistently across different values of $\sigma$ and always converges with thousands of simulation samples.}
On the contrary, the accuracy of the SIR-based approach is strongly dependent on the noise. 
{Particularly, its convergence is faster when $\sigma$ is large, and particularly slow when $\sigma$ is small (i.e., the observation is more informative).}
{These results demonstrate the robustness of our approach to measurement noise intensities, offering particular advantages when observations are informative.}

\section{Conclusion}

Reaction network identification is key to many biological studies.
However, current state-of-the-art methods suffer from computational inefficiency, as they rely on Monte Carlo techniques for rare-event estimation, which are known to be problematic.
This paper discovers, via simulation, an interesting phenomenon that for certain networks, the Fourier transform of the measurement data has a Gaussian-like distribution, thereby allowing for a novel identification algorithm that uses only mean and variance estimates.   
Since these estimates can be accurately obtained by a much reduced number of simulation samples, the proposed algorithm can be significantly more accurate than existing approaches. 
This improvement is also illustrated in a numerical example of a gene transcription system. 
From this numerical study, we also show that our method is robust to measurement noise intensities, making it particularly advantageous when this noise intensity is small, or equivalently, when the measurement data are more informative. 

Many topics deserve further investigation in future work. 
First, a theoretical analysis is required to justify the Gaussianity of the transformed data and to explain the computational improvement offered by this new approach. 
One of our forthcoming papers will address this topic to some extent. 
Second, our method can be combined with other advanced computational approaches, such as the control variates \citep{backenkohler2022variance}, marginal methods \citep{zechner2014scalable,fang2024advanced}, and Pad\'e approximations \citep{gupta2022frequency}, 
to further improve computational efficiency. 
Third, other transforms (i.e., the Laplace transform and wavelets) may exhibit properties similar to those of the Fourier transform observed in this study.
Therefore, it is worthwhile to investigate which transform is optimal in our framework to achieve the optimal computational efficiency and accuracy. 



\bibliography{ifacconf}

@article{jia2021frequency,
  title={Frequency domain analysis of fluctuations of mRNA and protein copy numbers within a cell lineage: theory and experimental validation},
  author={Jia, Chen and Grima, Ramon},
  journal={Physical Review X},
  volume={11},
  number={2},
  pages={021032},
  year={2021},
  publisher={APS}
}

@article{song2019frequency,
  title={Frequency spectrum of chemical fluctuation: A probe of reaction mechanism and dynamics},
  author={Song, Sanggeun and Yang, Gil-Suk and Park, Seong Jun and Hong, Sungguan and Kim, Ji-Hyun and Sung, Jaeyoung},
  journal={PLoS computational biology},
  volume={15},
  number={9},
  pages={e1007356},
  year={2019},
  publisher={Public Library of Science San Francisco, CA USA}
}

@article{heitzman2025practical,
  title={A practical identifiability criterion leveraging weak-form parameter estimation},
  author={Heitzman-Breen, Nora and Dukic, Vanja and Bortz, David M},
  journal={arXiv preprint arXiv:2506.17373},
  year={2025}
}

@article{messenger2024weak,
  title={Weak-form inference for hybrid dynamical systems in ecology},
  author={Messenger, Daniel and Dwyer, Greg and Dukic, Vanja},
  journal={Journal of the Royal Society Interface},
  volume={21},
  number={221},
  pages={20240376},
  year={2024},
  publisher={The Royal Society}
}

@article{gupta2022frequency,
  title={Frequency spectra and the color of cellular noise},
  author={Gupta, Ankit and Khammash, Mustafa},
  journal={Nature communications},
  volume={13},
  number={1},
  pages={4305},
  year={2022},
  publisher={Nature Publishing Group UK London}
}

@inproceedings{wildner2019moment,
  title={Moment-based variational inference for Markov jump processes},
  author={Wildner, Christian and Koeppl, Heinz},
  booktitle={International Conference on Machine Learning},
  pages={6766--6775},
  year={2019},
  organization={PMLR}
}

@article{ben2023learning,
  title={Learning-based importance sampling via stochastic optimal control for stochastic reaction networks},
  author={Ben Hammouda, Chiheb and Ben Rached, Nadhir and Tempone, Ra{\'u}l and Wiechert, Sophia},
  journal={Statistics and Computing},
  volume={33},
  number={3},
  pages={58},
  year={2023},
  publisher={Springer}
}

@article{golightly2019efficient,
  title={Efficient sampling of conditioned Markov jump processes},
  author={Golightly, Andrew and Sherlock, Chris},
  journal={Statistics and Computing},
  volume={29},
  number={5},
  pages={1149--1163},
  year={2019},
  publisher={Springer}
}

@incollection{backenkohler2022variance,
  title={Variance reduction in stochastic reaction networks using control variates},
  author={Backenk{\"o}hler, Michael and Bortolussi, Luca and Wolf, Verena},
  booktitle={Principles of Systems Design: Essays Dedicated to Thomas A. Henzinger on the Occasion of His 60th Birthday},
  pages={456--474},
  year={2022},
  publisher={Springer}
}

@article{zechner2014scalable,
  title={Scalable inference of heterogeneous reaction kinetics from pooled single-cell recordings},
  author={Zechner, Christoph and Unger, Michael and Pelet, Serge and Peter, Matthias and Koeppl, Heinz},
  journal={Nature methods},
  volume={11},
  number={2},
  pages={197--202},
  year={2014},
  publisher={Nature Publishing Group US New York}
}

@article{bronstein2015bayesian,
  title={Bayesian inference of reaction kinetics from single-cell recordings across a heterogeneous cell population},
  author={Bronstein, Leo and Zechner, Christoph and Koeppl, Heinz},
  journal={Methods},
  volume={85},
  pages={22--35},
  year={2015},
  publisher={Elsevier}
}

@article{fang2022stochastic,
  title={Stochastic filtering for multiscale stochastic reaction networks based on hybrid approximations},
  author={Fang, Zhou and Gupta, Ankit and Khammash, Mustafa},
  journal={Journal of Computational Physics},
  volume={467},
  pages={111441},
  year={2022},
  publisher={Elsevier}
}

@article{schnoerr2017approximation,
  title={Approximation and inference methods for stochastic biochemical kinetics—a tutorial review},
  author={Schnoerr, David and Sanguinetti, Guido and Grima, Ramon},
  journal={Journal of Physics A: Mathematical and Theoretical},
  volume={50},
  number={9},
  pages={093001},
  year={2017},
  publisher={IOP Publishing}
}

@article{stephens2003light,
  title={Light microscopy techniques for live cell imaging},
  author={Stephens, David J and Allan, Victoria J},
  journal={science},
  volume={300},
  number={5616},
  pages={82--86},
  year={2003},
  publisher={American Association for the Advancement of Science}
}

@book{asmussen2007stochastic,
  title={Stochastic simulation: algorithms and analysis},
  author={Asmussen, S{\o}ren and Glynn, Peter W},
  volume={57},
  year={2007},
  publisher={Springer}
}

@article{deligiannidis2018correlated,
  title={The correlated pseudomarginal method},
  author={Deligiannidis, George and Doucet, Arnaud and Pitt, Michael K},
  journal={Journal of the Royal Statistical Society Series B: Statistical Methodology},
  volume={80},
  number={5},
  pages={839--870},
  year={2018},
  publisher={Oxford University Press}
}

@article{chopin2013smc2,
  title={SMC2: an efficient algorithm for sequential analysis of state space models},
  author={Chopin, Nicolas and Jacob, Pierre E and Papaspiliopoulos, Omiros},
  journal={Journal of the Royal Statistical Society Series B: Statistical Methodology},
  volume={75},
  number={3},
  pages={397--426},
  year={2013},
  publisher={Oxford University Press}
}

@article{sisson2007sequential,
  title={Sequential monte carlo without likelihoods},
  author={Sisson, Scott A and Fan, Yanan and Tanaka, Mark M},
  journal={Proceedings of the National Academy of Sciences},
  volume={104},
  number={6},
  pages={1760--1765},
  year={2007},
  publisher={National Academy of Sciences}
}

@article{goodfellow2020generative,
  title={Generative adversarial networks},
  author={Goodfellow, Ian and Pouget-Abadie, Jean and Mirza, Mehdi and Xu, Bing and Warde-Farley, David and Ozair, Sherjil and Courville, Aaron and Bengio, Yoshua},
  journal={Communications of the ACM},
  volume={63},
  number={11},
  pages={139--144},
  year={2020},
  publisher={ACM New York, NY, USA}
}

@article{kobyzev2020normalizing,
  title={Normalizing flows: An introduction and review of current methods},
  author={Kobyzev, Ivan and Prince, Simon JD and Brubaker, Marcus A},
  journal={IEEE transactions on pattern analysis and machine intelligence},
  volume={43},
  number={11},
  pages={3964--3979},
  year={2020},
  publisher={IEEE}
}

@article{angus1994probability,
  title={The probability integral transform and related results},
  author={Angus, John E},
  journal={SIAM review},
  volume={36},
  number={4},
  pages={652--654},
  year={1994},
  publisher={SIAM}
}

@article{hastings1970monte,
  title={Monte Carlo sampling methods using Markov chains and their applications},
  author={Hastings, W Keith},
  year={1970},
  publisher={Oxford University Press}
}

@article{metropolis1953equation,
  title={Equation of state calculations by fast computing machines},
  author={Metropolis, Nicholas and Rosenbluth, Arianna W and Rosenbluth, Marshall N and Teller, Augusta H and Teller, Edward},
  journal={The journal of chemical physics},
  volume={21},
  number={6},
  pages={1087--1092},
  year={1953},
  publisher={American Institute of Physics}
}

@article{toni2009approximate,
  title={Approximate Bayesian computation scheme for parameter inference and model selection in dynamical systems},
  author={Toni, Tina and Welch, David and Strelkowa, Natalja and Ipsen, Andreas and Stumpf, Michael PH},
  journal={Journal of the Royal Society Interface},
  volume={6},
  number={31},
  pages={187--202},
  year={2009},
  publisher={The Royal Society London}
}

@article{fang2023convergence,
  title={Convergence of regularized particle filters for stochastic reaction networks},
  author={Fang, Zhou and Gupta, Ankit and Khammash, Mustafa},
  journal={SIAM Journal on Numerical Analysis},
  volume={61},
  number={2},
  pages={399--430},
  year={2023},
  publisher={SIAM}
}

@article{wang2020survey,
  title={A survey on Bayesian deep learning},
  author={Wang, Hao and Yeung, Dit-Yan},
  journal={ACM computing surveys (csur)},
  volume={53},
  number={5},
  pages={1--37},
  year={2020},
  publisher={ACM New York, NY, USA}
}

@article{andrieu2010particle,
  title={Particle markov chain monte carlo methods},
  author={Andrieu, Christophe and Doucet, Arnaud and Holenstein, Roman},
  journal={Journal of the Royal Statistical Society Series B: Statistical Methodology},
  volume={72},
  number={3},
  pages={269--342},
  year={2010},
  publisher={Oxford University Press}
}

@article{doucet2009tutorial,
	title={A tutorial on particle filtering and smoothing: Fifteen years later},
	author={Doucet, Arnaud and Johansen, Adam M and others},
	journal={Handbook of nonlinear filtering},
	volume={12},
	number={656-704},
	pages={3},
	year={2009}
}

@inproceedings{gordon1993novel,
	title={Novel approach to nonlinear/non-Gaussian Bayesian state estimation},
	author={Gordon, Neil J and Salmond, David J and Smith, Adrian FM},
	booktitle={IEE proceedings F (radar and signal processing)},
	volume={140},
	number={2},
	pages={107--113},
	year={1993},
	organization={IET}
}

@article{browning2020identifiability,
  title={Identifiability analysis for stochastic differential equation models in systems biology},
  author={Browning, Alexander P and Warne, David J and Burrage, Kevin and Baker, Ruth E and Simpson, Matthew J},
  journal={Journal of the Royal Society Interface},
  volume={17},
  number={173},
  pages={20200652},
  year={2020},
  publisher={The Royal Society}
}

@article{golightly2011bayesian,
  title={Bayesian parameter inference for stochastic biochemical network models using particle Markov chain Monte Carlo},
  author={Golightly, Andrew and Wilkinson, Darren J},
  journal={Interface focus},
  volume={1},
  number={6},
  pages={807--820},
  year={2011},
  publisher={The Royal Society}
}

@article{fang2024advanced,
  title={Advanced methods for gene network identification and noise decomposition from single-cell data},
  author={Fang, Zhou and Gupta, Ankit and Kumar, Sant and Khammash, Mustafa},
  journal={Nature Communications},
  volume={15},
  number={1},
  pages={4911},
  year={2024},
  publisher={Nature Publishing Group UK London}
}

@book{anderson2015stochastic,
  title={Stochastic analysis of biochemical systems},
  author={Anderson, David F and Kurtz, Thomas G},
  volume={674},
  year={2015},
  publisher={Springer}
}

\end{document}